# Multi-laser stabilization with an atomic-disciplined photonic integrated resonator

Andrei Isichenko[1], Andrew S. Hunter[1], Nitesh Chauhan[1], John R. Dickson[2], T. Nathan Nunley[3,4], Josiah R. Bingaman[2], David A. S. Heim[1], Mark W. Harrington[1], Kaikai Liu[1], Paul D. Kunz[2,4], and Daniel J. Blumenthal[1,*]

[1] *Department of Electrical and Computer Engineering, University of California Santa Barbara, Santa Barbara, California 93106, USA*
[2] *Department of Physics, Center of Complex Quantum Systems, The University of Texas at Austin, Austin, Texas 78712, USA*
[3] *General Technical Services, 1451 NJ-34, Wall Township, NJ 07727 USA*
[4] *DEVCOM Army Research Laboratory South, Austin, Texas 78712, USA*
*\*danb@ucsb.edu*

**Abstract:** Precision atomic and quantum experiments rely on ultra-stable narrow linewidth lasers constructed using table-top ultra-low expansion reference cavities. These experiments often require multiple lasers, operating at different wavelengths, to perform key steps used in state preparation and measurement required in quantum sensing and computing. This is traditionally achieved by disciplining a cavity-stabilized laser to a key atomic transition and then transferring the transition linewidth and stability to other lasers using the same reference cavity in combination with bulk-optic frequency shifting such as acousto-optic modulators. Transitioning such capabilities to a low cost photonic-integrated platform will enable a wide range of portable, low power, scalable quantum experiments and applications. Yet, today's bulk optic approaches pose challenges related to lack of cavity tunability, large free spectral range, and limited photonic integration potential. Here, we address these challenges with demonstration of an agile photonic-integrated 780 nm ultra-high-Q tunable silicon nitride reference cavity that performs multiple critical experimental steps including laser linewidth narrowing, high resolution rubidium spectroscopy, dual-stage stabilization to a rubidium transition, and stability transfer to other lasers. We achieve up to 20 dB of frequency noise reduction at 10 kHz offset, precision spectroscopy over a 250 MHz range, and dual-stage locking to rubidium with an Allan deviation of $8.5 \times 10^{-12}$ at 1 s and up to 40 dB reduction at 100 Hz. We further demonstrate the transfer of this atomic stability to a second laser, via the rubidium-disciplined cavity, and demonstrate multi-wavelength Rydberg electrometry quantum sensing. These results pave the path for integrated, compact, and scalable solutions for quantum sensing, computing and other atomic and trapped ion applications.

## 1. INTRODUCTION

Atomically disciplined lasers are critical for next-generation precision applications including quantum sensing [1], quantum computing [2,3], navigation [4], and timekeeping [5]. These applications require lasers with low frequency noise and stabilization to an atomic energy transition. The disciplining of lasers to an atom or trapped-ion enables applications such as optical atomic clocks [6], high-fidelity neutral atom sensors and computers [2,3] or trapped-ion qubits [7], and key supporting functions such as atomic spectroscopy [8] and qubit state preparation and measurement (SPAM) [7]. In particular, quantum applications such as Rydberg electrometry [9–11] require the simultaneous stabilization of multiple laser frequencies to atomic transitions or relative to a main transition. This is achieved by Pound-Drever-Hall

(PDH) locking each laser to one or more low-drift ultra-low expansion (ULE), high-finesse bulk-optic reference cavities [12], by using atomic spectroscopy in combination with a tunable or scanning cavity as a transfer cavity [13], or by using a fiber frequency comb system. In the latter approach, the transfer cavity is locked to an atom-referenced laser, and the resulting stability is transferred to additional lasers locked to the nearest cavity resonances. Tunable high-finesse transfer cavities (THFTCs) have also been demonstrated for linewidth narrowing, including stabilization of 780 nm and 960 nm lasers for single-atom coherent Rydberg excitations (~10 kHz integral linewidths, 22 kHz long-term drift) [14]. With ULE cavities, given the large FSR, frequency shifting is employed to fill the frequency gap between the available cavity modes and the desired atomic transitions using an electro-optic modulator (EOM) or an acousto-optic modulator (AOM) and offset sideband [15] or offset beat-note locking [16]. Tuning of the cavity reduces the need for bulk frequency shifters, however, table-top ULE reference cavities are not always tunable and do not lend easily to integration. The tunability of the reference cavity is therefore a key feature of this locking technique.

Recent advances in miniaturization of high quality factor (Q), high finesse reference cavities across multiple platforms [17–23] have provided progress towards these goals. Centimeter-scale microcavities [23] demonstrate short-term stability on the order of $10^{-13}$ at 1 second, yet, have limited tunability and slow tuning speeds on the order of minutes as well as large FSRs over 10 GHz. An important approach is the CMOS fabrication compatible integrated low loss silicon nitride platform to fabricate coil and spiral resonators [21,24]. These resonators have significantly lower FSR owing to their meter-scale cavity lengths [21] and can be dynamically tuned using on-chip stress-optic [25–27] and thermal [28] actuators. However, such integrated tunable cavities have not yet been demonstrated for atomic spectroscopy, multi-wavelength atomic transfer cavity locking, or for stability transfer in quantum applications.

Here we report a photonic-integrated approach to perform key steps used in the preparation and measurement of atomic states in quantum experiments. We demonstrate a tunable, integrated ultra-high-quality (UHQ) factor silicon nitride resonator cavity that simultaneously performs laser linewidth narrowing and frequency noise reduction, rubidium atomic spectroscopy, dual-stage laser locking to the rubidium $D_2$ transition, and functions as a multi-laser transfer cavity lock for a Rydberg RF electrometry demonstration. The cavity is a silicon nitride ring resonator with a Q factor of 130 million, finesse 982, and a thermal actuator that can continuously tune the resonance with an efficiency of 19 MHz/mW. We stabilize a 780 nm semiconductor laser via PDH locking to the integrated cavity and perform spectroscopy by actuating the thermal tuner and sweeping the linewidth narrowed 780 nm laser over the full rubidium hyperfine transition range. We then simultaneously stabilize the laser and integrated cavity to the rubidium $D_2$ line using a dual stage PDH lock, achieving up to 4 orders magnitude close-to-carrier frequency noise reduction, reduction of the laser integral linewidth from 5 MHz to 326 kHz, and measure stabilization of $8.5 \times 10^{-12}$ at 1 second -- a two order of magnitude improvement compared to the $9 \times 10^{-10}$ at 1 second by locking to the cavity only. Lastly, we use the transfer cavity locked to the rubidium $D_2$ transition to stabilize a 776 nm laser and demonstrate multi-wavelength Rydberg RF sensing. These results highlight the potential of photonic integrated tunable reference cavities to drive compact, scalable, and precise atomic and quantum systems.

## 2. RESONATOR FOR LASER LOCKING AND ATOMIC STABILIZATION

The working principle and functional stages are shown in Figure 1. A commercial single-frequency 780 nm laser (Photodigm™ DBR) is Pound-Drever-Hall (PDH) locked to the cavity resonance for linewidth narrowing and frequency noise reduction, forming the first locking stage (Fig. 1(a)). Thermal tuning of the cavity enables frequency control of the locked laser (Fig. 1(b)) and scanning the cavity allows for locating the rubidium spectroscopy peaks. In the second locking stage (Fig. 1(c)), the signal from the spectroscopy is fed back to the cavity tuner, stabilizing the PIC-locked laser to a hyperfine atomic transition and reducing the Allan deviation (ADEV) at longer averaging times. Lastly, the atom-stabilized resonator is used as a transfer cavity to lock multiple lasers to other cavity resonances, thereby transferring the stability from the reference laser to other lasers. In our work, this enables frequency stabilization of both 780 nm and 776 nm lasers for rubidium Rydberg electrometry, where the Autler-Townes splitting is used to detect radiofrequency (RF) electric fields as a quantum sensor.

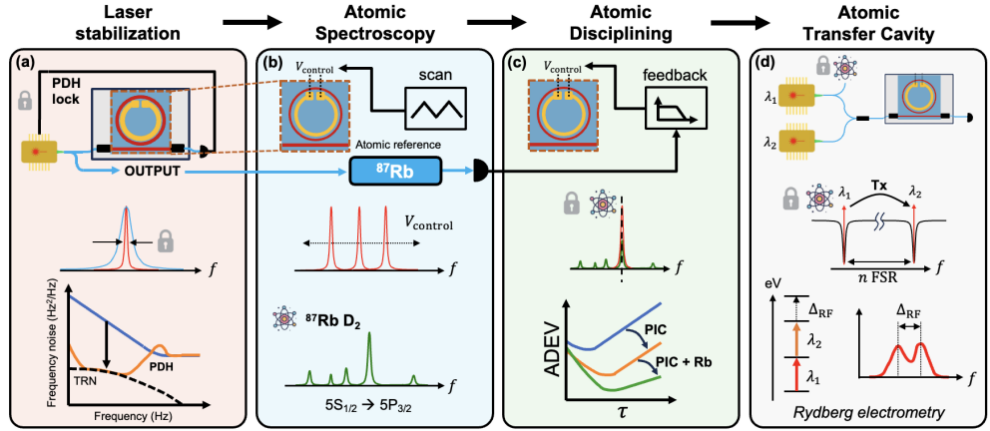

**Fig. 1.** Overview of the photonic-integrated ultra-high-Q (UHQ) resonator reference cavity and its application to atomic laser stabilization and multi-laser transfer cavity locking. (a) A 780 laser is Pound-Drever-Hall (PDH) locked to the integrated resonator, reducing its frequency noise to the cavity thermo-refractive noise limit (TRN) and causing the laser to closely track the resonator. (b) The same PIC resonator is thermally tuned causing the laser to scan across the rubidium D2 absorption spectrum. (c) Once aligned, feedback from the Rb spectroscopy signal is applied to the thermal tuner, implementing dual-stage locking. This disciplines the resonator frequency to the absolute atomic reference, improving long-term frequency stability as illustrated by the reduced Allan deviation (ADEV). (d) The stabilized resonator acts as a transfer cavity to lock multiple lasers, transferring the atomic stability to the second laser. This enables precise alignment of excitation lasers used in rubidium Rydberg electrometry, where a separate rubidium cell is used to detect RF signals via the measured Autler-Townes splitting ($\Delta_{RF}$).

### A. Resonator characterization

The integrated cavity consists of a fiber pigtailed, ultra-high-Q SiN ring resonator with a thermal tuner, fabricated on a 200 mm wafer from a CMOS foundry [29]. We characterize the resonator quality factor and tuning in Figure 2. The resonator intrinsic $Q_i$ = 130 M and loaded $Q_L$ = 69 M for the $TM_0$ mode are measured using an unbalanced Mach-Zehnder interferometer (MZI) [22,30] resulting in a linewidth 5.6 MHz and a loss of 0.44 dB/m (Fig. 2(a)). The ring resonator radius is 5.84 mm corresponding to a 5.5 GHz free-spectral range (FSR) and a finesse $\mathcal{F}$ = 982. We measure the relative resonance shift as a function of power applied to the thermal tuner as shown in Fig. 2(b) yielding a linear tuning coefficient of 19 MHz/mW. The measured tuning range of over 400 MHz is sufficient to cover the rubidium $D_2$ line spectroscopy peaks for SAS [31]. The device design and fabrication details are provided in Supplementary Information Note 1 and Figure S1.

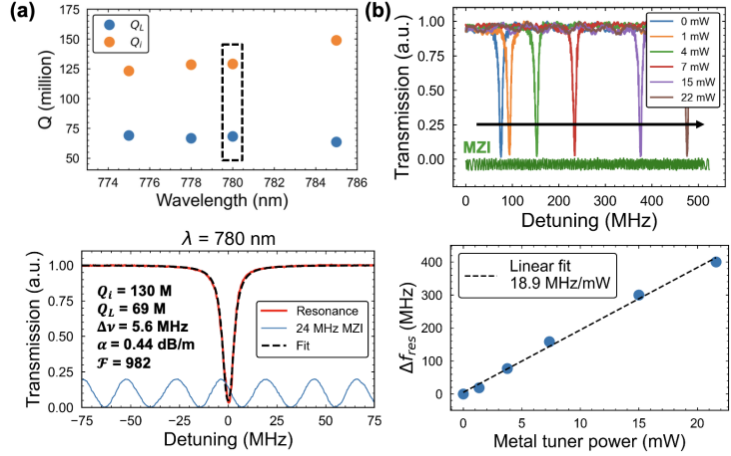

**Fig. 2**. Tunable integrated reference cavity characterization. (a) Intrinsic ($Q_i$) and loaded ($Q_L$) Q measurements for wavelengths near the rubidium 780 nm transition and transmission trace and fitting for a resonance at 780 nm achieving a loss $\alpha$ of 0.44 dB/m, total linewidth $\Delta\nu$ of 5.6 MHz, and finesse $\mathcal{F}$ of 982. The frequency detuning is calibrated with an unbalanced Mach-Zehnder interferometer (MZI, blue trace). (b) Static tuning of the thermal actuator shows a high extinction ratio across the measured >400 MHz tuning range. Linear fit to extract the 19 MHz/mW thermal tuning coefficient of the resonance shift $\Delta f_{\text{res}}$.

## B. Laser frequency stabilization

The integrated high-Q resonator cavity enables strong frequency noise reduction of a 780 nm semiconductor laser using PDH locking. We package the resonator with polarization-maintaining (PM) fibers and internal temperature control for stable, plug-and-play locking with single-frequency 780 nm lasers. In this experiment we used a 780 nm Distributed Bragg Reflector laser (DBR, Photodigm™) and a Velocity TLB-6730-P tunable external cavity diode laser (ECDL, Newport™) with tuning a range of 765 – 781 nm. We characterize the laser frequency noise using a fiber MZI optical frequency discriminator (OFD) [22].

We generate the error signal for PDH locking by monitoring the resonator transmission port fiber output on a photodetector and applying a 15.1 MHz modulation to either the laser current or by using a commercial electro-optic phase modulator (EOPM). The error signal is fed into a PID loop filter and the control signal is sent to the laser current control port (Fig. 3(a), top). The stabilized output is taken after a fiber splitter tap after the laser isolator. This PDH lock requires a minimum of ~2 mW going into the packaged reference cavity, allowing for 90% of the laser output available for subsequent experiments requiring frequency-stabilized light. We implement the lock with both the DBR and ECDL lasers which have different free-running fundamental (FLW) and integral linewidths (ILW). The OFD frequency noise (FN) measurement results for the PDH locking are shown in Fig. 3(b). At 10 kHz frequency offset, the PDH lock reduces the laser FN by 21 dB and 22 dB for the DBR and ECDL, respectively. The lock with the lower-noise ECDL laser achieves an ILW reduction from 23 kHz to 2 kHz and exhibits thermo-refractive noise (TRN) limited performance at 4 −30 kHz frequency offsets. The free-running and locked laser linewidths are summarized in Table 2 and we discuss the PDH loop details in Supplementary Information Note 2 and Fig. S2. In general, using a laser with a lower FLW and sufficient frequency modulation bandwidth enables tighter PDH locking and reduced ILW. Recent demonstrations of photonic-integrated external cavity tunable lasers (ECTLs) with agile tuning capabilities [27], as well as locking a low-noise ECTL to a low-TRN coil resonator [32], have achieved sub-kHz ILW performance in the C-band.

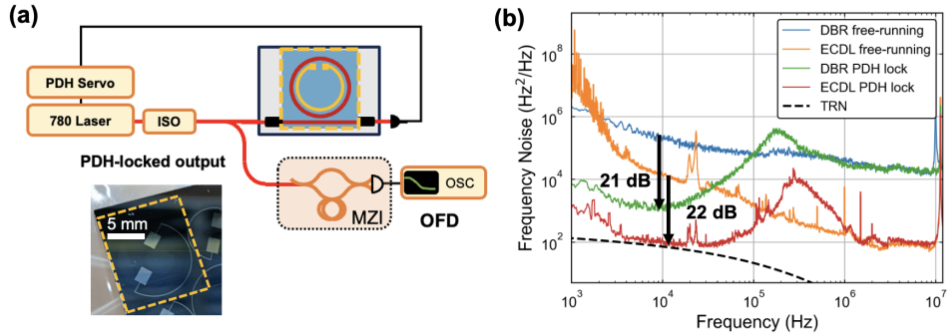

**Fig. 3**. Laser frequency stabilization to the tunable reference cavity. (a) Schematic for Pound-Drever-Hall (PDH) locking in the integrated resonator cavity ISO: optical isolator, OFD: optical frequency discriminator. Inset: photo of the metallized resonator cavity. (b) Laser frequency noise measurements with the OFD for the DBR laser and the ECDL laser. TRN: thermo-refractive noise.

Table 1. Laser linewidths achieved with packaged tunable PIC stabilization.

| Configuration | $1/\pi$ ILW | $\beta$-separation ILW | ILW range |
|---|---|---|---|
| DBR, free-running | 182 kHz [a] | 5.1 MHz | 0.1 Hz – 8 MHz |
| DBR, PIC lock | 246 kHz [a] | 306 kHz | 0.1 Hz – 8 MHz |
| ECDL, free-running | 23 kHz | 354 kHz | 1 kHz – 8 MHz |
| ECDL, PIC lock | 1.5 kHz | 2.3 kHz | 1 kHz – 8 MHz |
| PIC TRN-limited | 316 Hz | 470 Hz | 0.1 Hz – 8 MHz |

[a] The increase in the locked $1/\pi$ ILW for the DBR is due to the location of the servo bump, which arises from limited phase margin in the feedback loop. Combined with the high fundamental linewidth of the laser (33 kHz) and the limited PDH loop gain, this places stricter requirements on the ability of the servo to suppress mid-range frequency noise (see Supplementary Note 2).

### C. Stabilized laser tuning and modulation

Precise frequency tuning and dynamic modulation of the locked laser are essential for fine alignment and stabilization to an atomic reference. Here we demonstrate both frequency ramping for spectroscopy peak identification and dynamic modulation for generating the spectroscopy locking error signal. The approach to stabilizing the PDH-locked laser to rubidium spectroscopy is shown in in Fig. 4(a). We tap a portion of the laser light and send it through a free-space optical rubidium SAS module. We demonstrate locked laser sweeping by applying a 500 Hz voltage ramp to the PIC resonator integrated thermal actuator (Fig. 4(b)) and show that at different DC heater powers the frequency noise of the laser remains unchanged (Fig. S3(a), Supplementary Information Note 3). This shifts the cavity resonance, enabling a continuous frequency sweep of the locked laser. The sweep range of over 250 MHz used in this experiment is sufficient to identify the spectroscopic hyperfine peaks based on their relative strengths [31]. We also characterize the dynamic frequency modulation response for generating an error signal to lock the laser frequency precisely to the atomic transition. A small-signal voltage modulation is applied to the thermal actuator, and the frequency response of the locked laser is measured (Fig. S3(b)). While maintaining the lock, we measure the modulation bandwidth (defined at the 180° phase-lag frequency) to be $f_{180°}$ = 15 kHz (Fig. S3(c)).

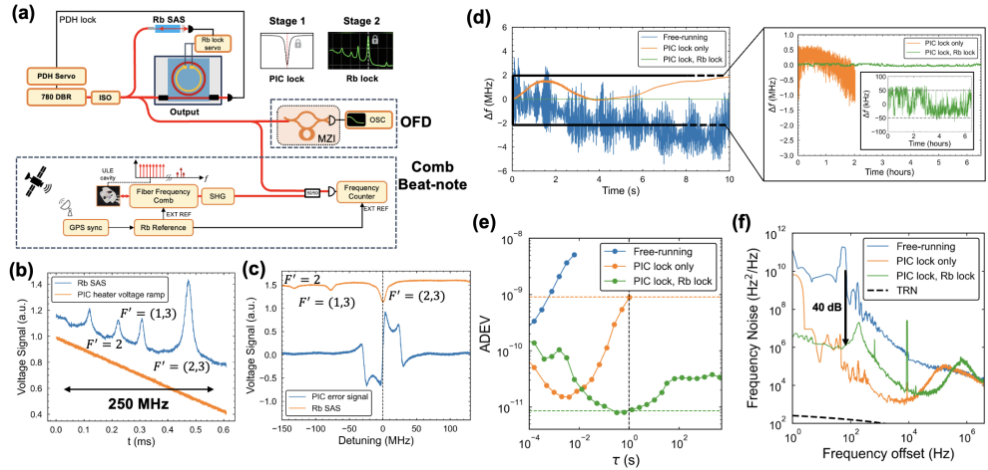

**Fig. 4.** Dual-stage locking to rubidium spectroscopy. (a) Experimental schematic for the dual-stage locking to the (stage 1) PIC cavity and to spectroscopy (stage 2). The full laser frequency noise and drift is characterized with the optical frequency discriminator (OFD) and ultra-low expansion (ULE) cavity-stabilized fiber frequency comb (Vescent FFC-100) beat-note system. SHG: second harmonic generation, EXT REF: external reference from the SRS FS725 Rb frequency standard. (b) Voltage ramp to the thermal tuner for scanning the locked laser across the rubidium $5\,S_{1/2}(F=2) \rightarrow 5\,P_{3/2}(F')$ hyperfine manifold. (c) Aligning PIC error signal and the target spectroscopy resonance. (d) Times series data for the beat-note frequency between frequency-doubled comb line and the dual-stage-locked laser. Inset: the dual-stage locked beat-note signal stays within 100 kHz over the 6 hours of measurement. (e) Allan deviation (ADEV) for different locking conditions: free-running laser, laser locked to the PIC only, laser locked to both PIC and Rb SAS. (f) Frequency noise stitched between the OFD and comb beat-note measurement methods.

### D. Rubidium-disciplined resonator

A dual-stage locking approach combines the short-term stability from the ultra-high-Q photonic integrated resonator cavity (PIC) with the long-term absolute frequency stability provided by rubidium spectroscopy (Fig. 4(a)). In the first stage, we PDH-lock a 780 nm DBR laser to the tunable integrated resonator for linewidth narrowing. Prior to engaging the lock, the cavity resonance error signal is tuned by setting the temperature setpoint of the packaged resonator, aligning the resonance to the rubidium $F' = (2,3)$ crossover transition (Fig. 4(c)). In the second stage, we lock the PIC cavity-stabilized laser to the rubidium SAS signal for long-term frequency referencing. To generate the necessary spectroscopy error signal for atomic locking, we apply a small-signal modulation (9 kHz) to the PIC thermal actuator. This spectroscopy-generated error signal is then fed back to the cavity thermal tuner via a Red Pitaya-based servo loop, to maintain the cavity resonance frequency for robust, long-term alignment with the atomic transition.

We next characterize the dual-stage locked laser performance using a combination of the OFD frequency noise system and a cavity-stabilized self-reference optical frequency comb (Fig. 4(a)). The frequency comb (Vescent FFC-100) is disciplined to a Rb frequency standard referenced to a GPS receiver. The comb system generates a beat-note between a frequency-doubled comb line at 780 nm and the locked laser and the beat-note is recorded on a frequency counter. The OFD provides short-term frequency noise at > 1 kHz offset frequencies and the comb beat-note is used for < 1 kHz offsets long-term drift characterization. The resulting frequency stability time series, Allan deviation (ADEV), and OFD/comb stitched frequency noise data is shown in Fig. 4(d-f) for three different configurations: the free-running DBR laser, the DBR laser locked only to the PIC, and the DBR laser under dual-stage (PIC and rubidium spectroscopy) lock.

The PIC-only lock reduces the short-term laser noise and stabilizes the drift to within several MHz, reducing the free-running laser ADEV by more than two orders of magnitude at ~millisecond timescales (Fig. 4 (d,e)). However, due to inherent cavity drift, PIC-only locking is insufficient for atomic experiments requiring stable absolute frequency referencing at averaging times greater than ~10 ms. The second-stage rubidium spectroscopy lock significantly improves the long-term stability, achieving fractional frequency stability of $8.5 \times 10^{-12}$ at 1 second averaging time -- a two-order-of-magnitude improvement compared to the PIC-only lock performance of $9 \times 10^{-10}$ at 1 second (Fig. 4(e)). The dual-stage lock confines the laser frequency drift to within ±50 kHz over 6 hours of continuous operation (Fig. 4(d), inset).

The frequency noise of the locked laser is shown in Fig. 4(f), with data from two different measurement systems (comb beat note and OFD) stitched together. At frequency offsets 100 Hz – 10 kHz the laser frequency noise increases due to thermal tuner feedback loop in the rubidium SAS lock. The β-separation integral linewidth of the free-running 780 nm laser is reduced from 5 MHz to 306 kHz for the PIC lock only and to 326 kHz for the dual-stage lock. The slight increase in integral linewidth from 306 kHz (PIC-only lock) to 326 kHz (dual-stage lock) is due in part due to the limited bandwidth of the thermal actuator. Future investigations will focus on enhancing actuator performance, such as addressing nonlinearities and hysteresis, and quantifying systematic frequency shifts arising from rubidium cell temperature and probe beam power fluctuations to further improve the long-term stability. We summarize the calculated ILWs and ADEV for the different lock loops in Table 2.

**Table 2.** Laser linewidths and ADEV with the dual-stage locked tunable integrated cavity.

| DBR Laser Lock | $1/\pi$ ILW | $\beta$-separation ILW | ILW range | ADEV at 1 ms / 1 s / 10 s |
|---|---|---|---|---|
| Free-running | 182 kHz [a] | 5.1 MHz | 0.1 Hz – 8 MHz | $1.4 \times 10^{-9}$ / - / - |
| PIC lock only | 246 kHz [a] | 306 kHz | 0.1 Hz – 8 MHz | $2 \times 10^{-11}$ / $9 \times 10^{-10}$ / - |
| PIC and Rb lock | 346 kHz | 326 kHz | 0.1 Hz – 8 MHz | $8 \times 10^{-11}$ / $8.5 \times 10^{-12}$ / $1.5 \times 10^{-11}$ |

[a] See footnote in Table 1.

## 3. MULTI-WAVELENGTH RYDBERG SENSING

We demonstrate the use of our tunable integrated resonator to transfer rubidium stability to two lasers in a three-wavelength Rydberg electrometry, where three distinct wavelengths are used to sequentially drive transitions from a ground state to an intermediate excited state and subsequently to a high-lying Rydberg state. Rydberg electrometry allows for measuring RF electric fields with high sensitivity across a wide range of frequencies using room-temperature rubidium vapor cells for precision sensing, RF field mapping, and antenna calibration [11]. The energy diagram for the 780 nm (probe laser), 776 nm (dressing laser), and 1270 nm (coupling laser) is shown in Fig. 5(a). In this application, the long-term stability of the probe and dressing lasers are important to achieving high sensitivity for the RF electric field measurements.

The tunable integrated resonator is used to lock both the 780 nm and 776 nm lasers using a single rubidium spectroscopy reference (Fig. 5(b)). The resonator is critically coupled with high Q resonances over the 776–785 nm range, providing a strong PDH error signal slope for stabilization. We first stabilize a 780 ECDL to the transfer cavity resonator and rubidium spectroscopy using a dual-stage lock (described in the previous section). Next, we stabilize the 776 ECDL to the PIC resonator using electro-optic phase modulator (EOPM) sideband locking, thereby transferring the rubidium-stabilized frequency stability from the 780 nm laser to the 776 nm laser. The details of the lock loop are described in Supplementary Information Note 4.

Finally, the 1270 nm Rydberg coupling laser is scanned through the expected Rydberg transition frequency, while the transmission of the 780 nm probe beam is monitored to detect electromagnetically induced transparency (EIT) signals corresponding to Rydberg excitation (Fig. 5(c)). With a 2.76 GHz RF field incident on the atoms, we observe clear Autler-Townes splitting of the Rydberg resonance, from which the RF electric field amplitude can be derived.

The Rydberg EIT signal (RF off) from the 1270 laser transmission scan is continuously sampled for 20 minutes (Fig. 5(d)). We quantify the signal quality from the trace as a function of time for three different locking conditions: both 780 nm and 776 nm lasers free-running, both lasers locked to the PIC transfer cavity, and both lasers locked to a table-top vacuum-gap ultra-low expansion (ULE) quad-bore reference cavity. The details of the EIT signal data processing are discussed in Supplementary Information Note 4. We show that locking both lasers to the PIC-based transfer cavity significantly enhances the signal stability relative to the free-running condition and achieves comparable performance to the ULE quad-bore cavity. To further support these findings, we evaluated the frequency stability of the PIC-based transfer cavity lock using the comb beat-note measurement system (shown in Fig. 3(a)), as discussed in Supplementary Information Note 4. The corresponding ADEV traces (Fig. S4(b)) show that both lasers exhibit sub-$10^{-10}$ stability at 1–100 ms averaging times when locked to the PIC. At longer averaging times, the addition of rubidium spectroscopy feedback to the 780 nm laser further disciplines the cavity, reducing the ADEV of the transfer-locked 776 nm laser by nearly an order of magnitude at $\tau = 0.1$ s. The frequency noise at 1 kHz offset was reduced by 20 dB (Fig. S4(c)).

We also confirm that the same PIC resonator supports a 2.2 M loaded Q TE resonance at 1270 nm—used for Rydberg state coupling—suggesting that this wavelength could also be stabilized using the PIC (Fig. S1(d)). While the 1270 nm laser was scanned rather than locked in this work, future implementations could benefit from fully stabilizing all three lasers, especially when the 1270 nm beam serves as a fixed-frequency probe or sensing reference. These results validate the integrated transfer cavity approach for robust multi-wavelength stabilization, demonstrating its suitability for high-precision atomic and quantum experiments. The PIC-based cavity provides a compact and scalable approach to relative laser frequency control and reduces the need for separate atomic references for each laser, expensive fiber frequency combs, or table-top ULE reference cavities.

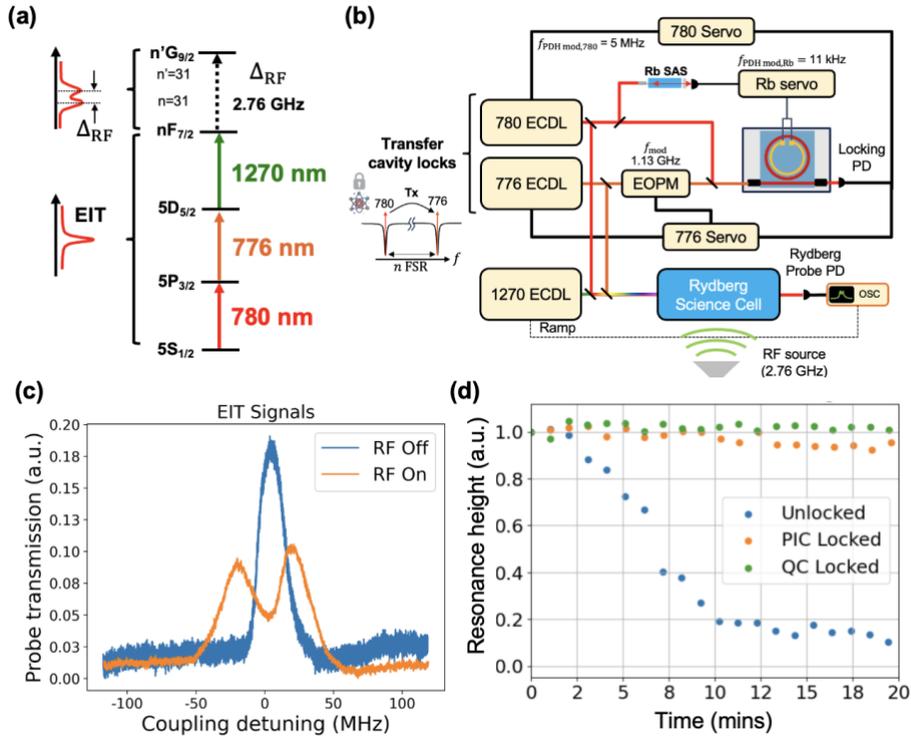

**Fig. 5**. Simultaneous multi-wavelength locking to PIC reference for three-wavelength Rydberg electrometry for RF field sensing. (a) Energy level diagram for three-wavelength transition. The alignment of all three optical transitions creates an electromagnetically-induced transparent (EIT) signal which is observed by measuring the 780 nm probe laser transmission. (b) Experimental setup for stabilizing the 780 and 776 external cavity diode lasers (ECDL) to the PIC transfer cavity. The 1270 nm coupling laser is scanned through the $|d\rangle \rightarrow |r\rangle$ $31F_{7/2}$ transition and the 780 nm probe transmission is recorded on a photodetector (PD). The 780 ECDL serves as the SAS-locked reference and the 776 ECDL PDH is PDH sideband-locked to a cavity resonance using an electro-optic phase modulator (EOPM). The external RF source incident at the atoms is operated at 2.76 GHz. (c) 780 probe laser transmission trace as a function of the 1270 laser scan. Strong Autler-Townes splitting is observed when the RF field is on. (d) Rydberg laser transmission signal as a function of time for 780 and 776 lasers free-running (blue), both locked to the Rb-disciplined PIC transfer cavity (orange), and both locked to a table-top quad-bore ULE reference cavity (green).

## 4. DISCUSSION AND CONCLUSION

In this work we present a photonic-integrated approach to perform key steps used in the preparation and measurement of atomic states in quantum experiments. These steps include spectroscopy of the atomic hyperfine structure, stabilizing a primary laser and a reference cavity to a desired atom transition, transferring the stability of the atom to additional laser frequencies using additional cavity resonances, performing state preparation of the atom population using the atomically stabilized set of lasers, and making a quantum measurement using a cavity referenced laser. Using a thermally actuated CMOS-foundry-fabricated tunable silicon nitride resonator, we realize a tunable locked laser source with 22 dB of frequency noise reduction at 10 kHz offset and scanning rubidium spectroscopy using precise control of the locked low noise laser frequency. After spectroscopy sweeping, we are able to choose the desired hyperfine transition and implement a dual-stage lock that combines the short-term stability of the integrated resonator with the long-term stability of the rubidium $D_2$ transition, achieving a fractional frequency stability of $8.5\times10^{-12}$ at 1 second and maintaining frequency drift to within 100 kHz for over 6 hours of continuous operation. We then use the same resonator to transfer the rubidium stability to a second laser and perform quantum atomic state preparation enabling electromagnetic field measurement in a three-wave Rydberg RF sensing experiment. The ability to successfully perform EIT detection on the probe light demonstrates that the rubidium-disciplined cavity resonances successfully transfers atomic stability to the second laser.

Looking forward, this approach can be applied to wide range of neutral atom and trapped-ion experiments that involve broader wavelength spectroscopy and coverage, narrower linewidth atomic transitions, and precision frequency metrology. Examples include probing the rubidium two-photon transition for optical atomic clocks [33,34], trapped ion qubits [35], and multi-wavelength stabilization for polyatomic molecule experiments with many repump wavelengths [36]. While the Rydberg electrometry was performed with a warm vapor cell, a similar stabilization architecture can be extended to cold-atom Rydberg architectures that require narrower laser linewidths [14]. Future work will also focus on precisely quantifying the transfer of frequency noise and stability in the multi-laser stabilization across a broader range of resonances and using longer cavities with smaller FSR.

The atomic-disciplined reference and transfer cavity operation offers a compact alternative to traditional table-top multi-bore ULE cavities, costly frequency comb systems, and reduces the need to use multiple atomic references. The agile integrated cavity overcomes limitations of table-top bulk-optic references, namely lack of agility, large free spectral range, and difficulty in planar photonic integration. Beyond the compact footprint and linewidth narrowing capabilities, photonic integrated resonators provide a scalable solution to support significantly lower FSR designs; for example, meter-scale coil resonators on a chip can provide FSRs in the range 10 - 100 MHz [21,37,38]. These low-FSR resonators increase lock-point density to help bridge the frequency gap between cavity resonances and atomic transitions without using power-consuming AOMs. This capability that would be challenging to realize with bulk-optic cavities that would be difficult to fabricate, thermally tune, and stabilize at comparable cavity length dimensions. Moreover, the integration of stress-optic tuning actuators, such as PZT-on-SiN, enables >10 MHz bandwidth actuation [25–27] provides fast, chip-scale tuning, overcoming the kHz-level bandwidth limits imposed by mechanical resonances in traditional table-top PZT-actuated cavities, and providing low power (order 10s nW) actuation companion to thermal tuning. Further, an integrated cavity with fast, continuous broadband tuning reduces the need for external frequency shifters such as AOMs, which are power consuming and typically limited to a few hundred MHz of tuning range.

Ultimately, co-designing photonic reference cavities with lasers [22,32,39–41], modulators [25,26,42] and atomic-photonic interfaces will enable fully chip-integrated quantum systems. This approach can be integrated with on-chip beam delivery, including grating emitters directly coupled to atomic cells for probing and trapping atoms [43,44] to further reduce system size and complexity. These advances highlight the potential of integrated photonic reference cavities as a scalable platform for neutral atom and trapped-ion quantum computing and sensing, ultimately enabling wafer-scale optical systems for timekeeping, navigation, and other quantum technologies.

**Funding.** This work is supported by the U.S. Army Research Laboratory (W911NF-22-2-0056), NASA Quantum Pathways Institute (80NSSC23K1343), and the NSF Q-SEnSE QLCI (OMA-2016244). This material is based upon work supported by, or in part by, the U. S. Army Research Laboratory and the U. S. Army Research Office under contract/grant number W911NF2310179. The views and conclusions contained in this document are those of the authors and should not be interpreted as representing the official policies, either expressed or implied, of the Army Research Laboratory or the U.S. Government. Any opinions, findings, and conclusions or recommendations expressed in this material are those of the author(s) and do not necessarily reflect the views of the National Aeronautics and Space Administration (NASA).

**Acknowledgment.** We thank Karl D. Nelson at Honeywell for the CMOS fabrication and Jiawei Wang at UCSB for the device metallization. The authors would also like to thank Matthew Hummon and John Kitching at NIST, Joseph Britton and Wance Wang at University of Maryland, and Joshua Hill, David Meyer, and Nathan O'Malley at DEVCOM Army Research Laboratory for useful discussions. We acknowledge Vescent Technologies for their help with setting up the optical frequency comb.

**Disclosures.** DJB: ColdQuanta D.B.A. Infleqtion (F, I, C, P). All other authors declare that there are no conflicts of interest related to this article.

**Data availability.** Data underlying the results presented in this paper are not publicly available at this time but may be obtained from the authors upon reasonable request.

**Supplemental document.** See Supplement 1 for supporting content.

**References**

1. B. Stray, A. Lamb, A. Kaushik, J. Vovrosh, A. Rodgers, J. Winch, F. Hayati, D. Boddice, A. Stabrawa, A. Niggebaum, M. Langlois, Y.-H. Lien, S. Lellouch, S. Roshanmanesh, K. Ridley, G. de Villiers, G. Brown, T. Cross, G. Tuckwell, A. Faramarzi, N. Metje, K. Bongs, and M. Holynski, "Quantum sensing for gravity cartography," Nature **602**(7898), 590–594 (2022).
2. X. Jiang, J. Scott, M. Friesen, and M. Saffman, "Sensitivity of quantum gate fidelity to laser phase and intensity noise," Phys. Rev. A **107**(4), 042611 (2023).
3. T. M. Graham, Y. Song, J. Scott, C. Poole, L. Phuttitarn, K. Jooya, P. Eichler, X. Jiang, A. Marra, B. Grinkemeyer, M. Kwon, M. Ebert, J. Cherek, M. T. Lichtman, M. Gillette, J. Gilbert, D. Bowman, T. Ballance, C. Campbell, E. D. Dahl, O. Crawford, N. S. Blunt, B. Rogers, T. Noel, and M. Saffman, "Multi-qubit entanglement and algorithms on a neutral-atom quantum computer," Nature **604**(7906), 457–462 (2022).
4. J. Kitching, "Chip-scale atomic devices," Applied Physics Reviews **5**(3), 031302 (2018).
5. A. D. Ludlow, M. M. Boyd, J. Ye, E. Peik, and P. O. Schmidt, "Optical atomic clocks," Rev. Mod. Phys. **87**(2), 637–701 (2015).
6. C. Audoin, V. Candelier, and N. Diamarcq, "A limit to the frequency stability of passive frequency standards due to an intermodulation effect," IEEE Transactions on Instrumentation and Measurement **40**(2), 121–125 (1991).
7. C. D. Bruzewicz, J. Chiaverini, R. McConnell, and J. M. Sage, "Trapped-ion quantum computing: Progress and challenges," Applied Physics Reviews **6**(2), 021314 (2019).
8. P. O. Schmidt, T. Rosenband, C. Langer, W. M. Itano, J. C. Bergquist, and D. J. Wineland, "Spectroscopy Using Quantum Logic," Science **309**(5735), 749–752 (2005).
9. J. A. Sedlacek, A. Schwettmann, H. Kübler, R. Löw, T. Pfau, and J. P. Shaffer, "Microwave electrometry with Rydberg atoms in a vapour cell using bright atomic resonances," Nature Phys **8**(11), 819–824 (2012).
10. N. Thaicharoen, K. R. Moore, D. A. Anderson, R. C. Powel, E. Peterson, and G. Raithel, "Electromagnetically induced transparency, absorption, and microwave-field sensing in a Rb vapor cell with a three-color all-infrared laser system," Phys. Rev. A **100**(6), 063427 (2019).
11. D. H. Meyer, Z. A. Castillo, K. C. Cox, and P. D. Kunz, "Assessment of Rydberg atoms for wideband electric field sensing," J. Phys. B: At. Mol. Opt. Phys. **53**(3), 034001 (2020).
12. D. G. Matei, T. Legero, S. Häfner, C. Grebing, R. Weyrich, W. Zhang, L. Sonderhouse, J. M. Robinson, J. Ye, F. Riehle, and U. Sterr, "1.5 μ m Lasers with Sub-10 mHz Linewidth," Phys. Rev. Lett. **118**(26), 263202 (2017).
13. E. Pultinevicius, M. Rockenhäuser, F. Kogel, P. Groß, T. Garg, O. E. Prochnow, and T. Langen, "A scalable scanning transfer cavity laser stabilization scheme based on the Red Pitaya STEMlab platform," Review of Scientific Instruments **94**(10), 103004 (2023).
14. Y. Zeng, K.-P. Wang, Y.-Y. Liu, X.-D. He, M. Liu, P. Xu, J. Wang, and M.-S. Zhan, "Stabilizing dual laser with a tunable high-finesse transfer cavity for single-atom Rydberg excitation," J. Opt. Soc. Am. B, JOSAB **35**(2), 454–459 (2018).
15. J. I. Thorpe, K. Numata, and J. Livas, "Laser frequency stabilization and control through offset sideband locking to optical cavities," Opt. Express, OE **16**(20), 15980–15990 (2008).
16. M. Mäusezahl, F. Munkes, and R. Loew, "Tutorial on laser locking techniques and the manufacturing of vapor cells for spectroscopy," New J. Phys. (2024).
17. Y.-H. Lai, D. Eliyahu, S. Ganji, R. Moss, I. Solomatine, E. Lopez, E. Tran, A. Savchenkov, A. Matsko, and S. Williams, "780 nm narrow-linewidth self-injection-locked WGM lasers," in *Laser Resonators, Microresonators, and Beam Control XXII* (SPIE, 2020), **11266**, pp. 78–84.
18. W. Loh, M. T. Hummon, H. F. Leopardi, T. M. Fortier, F. Quinlan, J. Kitching, S. B. Papp, and S. A. Diddams, "Microresonator Brillouin laser stabilization using a microfabricated rubidium cell," Opt. Express **24**(13), 14513 (2016).

19. W. Zhang, L. Stern, D. Carlson, D. Bopp, Z. Newman, S. Kang, J. Kitching, and S. B. Papp, "Ultranarrow Linewidth Photonic-Atomic Laser," Laser & Photonics Reviews **14**(4), 1900293 (2020).
20. W. Jin, Q.-F. Yang, L. Chang, B. Shen, H. Wang, M. A. Leal, L. Wu, M. Gao, A. Feshali, M. Paniccia, K. J. Vahala, and J. E. Bowers, "Hertz-linewidth semiconductor lasers using CMOS-ready ultra-high-Q microresonators," Nat. Photonics **15**(5), 346–353 (2021).
21. K. Liu, N. Chauhan, J. Wang, A. Isichenko, G. M. Brodnik, P. A. Morton, R. O. Behunin, R. O. Behunin, S. B. Papp, S. B. Papp, and D. J. Blumenthal, "36 Hz integral linewidth laser based on a photonic integrated 4.0 m coil resonator," Optica, OPTICA **9**(7), 770–775 (2022).
22. N. Chauhan, A. Isichenko, K. Liu, J. Wang, Q. Zhao, R. O. Behunin, P. T. Rakich, A. M. Jayich, C. Fertig, C. W. Hoyt, and D. J. Blumenthal, "Visible light photonic integrated Brillouin laser," Nat Commun **12**(1), 4685 (2021).
23. N. Jin, C. A. McLemore, D. Mason, J. P. Hendrie, Y. Luo, M. L. Kelleher, P. Kharel, F. Quinlan, S. A. Diddams, and P. T. Rakich, "Micro-fabricated mirrors with finesse exceeding one million," Optica, OPTICA **9**(9), 965–970 (2022).
24. L. Cheng, M. Zhao, Y. He, Y. Zhang, R. Meade, K. Vahala, M. Zhang, and J. Li, "Spiral resonator referenced low noise microwave generation via integrated optical frequency division," Photon. Res., PRJ **13**(7), 1991–1996 (2025).
25. J. Wang, K. Liu, M. W. Harrington, R. Q. Rudy, and D. J. Blumenthal, "Silicon nitride stress-optic microresonator modulator for optical control applications," Opt. Express, OE **30**(18), 31816–31827 (2022).
26. P. R. Stanfield, A. J. Leenheer, C. P. Michael, R. Sims, and M. Eichenfield, "CMOS-compatible, piezo-optomechanically tunable photonics for visible wavelengths and cryogenic temperatures," Opt. Express, OE **27**(20), 28588–28605 (2019).
27. G. Lihachev, A. Bancora, V. Snigirev, H. Tian, J. Riemensberger, V. Shadymov, A. Siddharth, A. Attanasio, R. N. Wang, D. A. Visani, A. Voloshin, S. A. Bhave, and T. J. Kippenberg, "Frequency agile photonic integrated external cavity laser," APL Photonics **9**(12), 126102 (2024).
28. A. Isichenko, N. Chauhan, J. Wang, M. W. Harrington, K. Liu, and D. J. Blumenthal, "Tunable Integrated 118 Million Q Reference Cavity for 780 nm Laser Stabilization and Rubidium Spectroscopy," in *CLEO 2023 (2023), Paper SF3K.4* (Optica Publishing Group, 2023), p. SF3K.4.
29. D. J. Blumenthal, R. Heideman, D. Geuzebroek, A. Leinse, and C. Roeloffzen, "Silicon Nitride in Silicon Photonics," Proc. IEEE **106**(12), 2209–2231 (2018).
30. S. Gundavarapu, G. M. Brodnik, M. Puckett, T. Huffman, D. Bose, R. Behunin, J. Wu, T. Qiu, C. Pinho, N. Chauhan, J. Nohava, P. T. Rakich, K. D. Nelson, M. Salit, and D. J. Blumenthal, "Sub-hertz fundamental linewidth photonic integrated Brillouin laser," Nature Photon **13**(1), 60–67 (2019).
31. D. A. Steck, "Rubidium 87 D Line Data," http://steck.us/alkalidata.
32. D. A. S. Heim, D. Bose, K. Liu, A. Isichenko, and D. J. Blumenthal, "Hybrid integrated ultra-low linewidth coil stabilized isolator-free widely tunable external cavity laser," Nat Commun **16**(1), 5944 (2025).
33. A. Isichenko, A. Kortyna, N. Chauhan, J. Wang, M. W. Harrington, J. Olson, and D. J. Blumenthal, "Tunable 778 nm Integrated Brillouin Laser Probe for a Rubidium Two-Photon Optical Atomic Clock," in *CLEO 2024 (2024), Paper SM1R.7* (Optica Publishing Group, 2024), p. SM1R.7.
34. K. W. Martin, G. Phelps, N. D. Lemke, M. S. Bigelow, B. Stuhl, M. Wojcik, M. Holt, I. Coddington, M. W. Bishop, and J. H. Burke, "Compact Optical Atomic Clock Based on a Two-Photon Transition in Rubidium," Phys. Rev. Applied **9**(1), 014019 (2018).
35. J.-R. Chen, T.-H. Suen, C.-Y. Kung, L.-B. Wang, and Y.-W. Liu, "High stability multiple-frequency cavity locking based on Doppler-free optogalvanic Calcium ion spectroscopy," Opt. Express **30**(15), 28170 (2022).
36. N. R. Hutzler, "Polyatomic molecules as quantum sensors for fundamental physics," Quantum Sci. Technol. **5**(4), 044011 (2020).
37. N. Chauhan, K. Liu, A. Isichenko, J. Wang, H. Timmers, and D. J. Blumenthal, "Integrated 3.0 meter coil resonator for λ = 674 nm laser stabilization," in *Frontiers in Optics + Laser Science 2022 (FIO, LS) (2022), Paper FM1E.1* (Optica Publishing Group, 2022), p. FM1E.1.
38. B. Li, W. Jin, L. Wu, L. Chang, H. Wang, B. Shen, Z. Yuan, A. Feshali, M. Paniccia, K. J. Vahala, and J. E. Bowers, "Reaching fiber-laser coherence in integrated photonics," Opt. Lett., OL **46**(20), 5201–5204 (2021).
39. J. P. McGilligan, K. Gallacher, P. F. Griffin, D. J. Paul, A. S. Arnold, and E. Riis, "Micro-fabricated components for cold atom sensors," Review of Scientific Instruments **93**(9), 091101 (2022).
40. A. Siddharth, A. Attanasio, S. Bianconi, G. Lihachev, J. Zhang, Z. Qiu, A. Bancora, S. Kenning, R. N. Wang, A. S. Voloshin, S. A. Bhave, J. Riemensberger, and T. J. Kippenberg, "Piezoelectrically tunable, narrow linewidth photonic integrated extended-DBR lasers," Optica, OPTICA **11**(8), 1062–1069 (2024).
41. A. Isichenko, A. S. Hunter, D. Bose, N. Chauhan, M. Song, K. Liu, M. W. Harrington, and D. J. Blumenthal, "Sub-Hz fundamental, sub-kHz integral linewidth self-injection locked 780 nm hybrid integrated laser," Sci Rep **14**(1), 27015 (2024).
42. D. Renaud, D. R. Assumpcao, G. Joe, A. Shams-Ansari, D. Zhu, Y. Hu, N. Sinclair, and M. Loncar, "Sub-1 Volt and high-bandwidth visible to near-infrared electro-optic modulators," Nat Commun **14**(1), 1496 (2023).
43. M. T. Hummon, S. Kang, D. G. Bopp, Q. Li, D. A. Westly, S. Kim, C. Fredrick, S. A. Diddams, K. Srinivasan, V. Aksyuk, and J. E. Kitching, "Photonic chip for laser stabilization to an atomic vapor with $10^{-11}$ instability," Optica, OPTICA **5**(4), 443–449 (2018).

44. A. Isichenko, N. Chauhan, D. Bose, J. Wang, P. D. Kunz, and D. J. Blumenthal, "Photonic integrated beam delivery for a rubidium 3D magneto-optical trap," Nat Commun **14**(1), 3080 (2023).

# MULTI-LASER STABILIZATION WITH AN ATOMIC-DISCIPLINED PHOTONIC INTEGRATED RESONATOR: SUPPLEMENTAL DOCUMENT

ANDREI ISICHENKO[1], ANDREW S. HUNTER[1], NITESH CHAUHAN[1], JOHN R. DICKSON[2], T. NATHAN NUNLEY[3,4], JOSIAH R. BINGAMAN[2], DAVID A. S. HEIM[1], MARK W. HARRINGTON[1], KAIKAI LIU[1], PAUL D. KUNZ[2,3] AND DANIEL J. BLUMENTHAL[1,*]

[1] Department of Electrical and Computer Engineering, University of California Santa Barbara, Santa Barbara, California 93106, USA
[2] Department of Physics, Center of Complex Quantum Systems, The University of Texas at Austin, Austin, Texas 78712, USA
[3] General Technical Services, 1451 NJ-34, Wall Township, NJ 07727 USA
[4] DEVCOM Army Research Laboratory South, Austin, Texas 78712, USA
*danb@ucsb.edu

## 1. Resonator design and fabrication

The ring resonator device consists of a $Si_3N_4$ waveguide with a 15 $\mu$m $SiO_2$ lower cladding, a 40 nm thick and 4 $\mu$m wide $Si_3N_4$ core, and a 6 $\mu$m $SiO_2$ upper cladding (Fig. S1(a)). The fabrication of these resonators includes the deposition of a 40 nm thick stoichiometric $Si_3N_4$ film using low-pressure chemical vapor deposition (LPCVD) on a 15 $\mu$m thermal oxide pre-grown on a 200 mm silicon wafer. Following the deep-ultraviolet (DUV) stepper lithography and etching, a 6 $\mu$m upper oxide cladding is deposited using tetraethoxysilane pre-cursor plasma-enhanced chemical vapor deposition (TEOS-PECVD) and the wafer is annealed at ~1100 deg C. After the wafer was diced, several die each containing four ring resonators were metallized with platinum electrodes of thickness 250 nm and width 100 um.

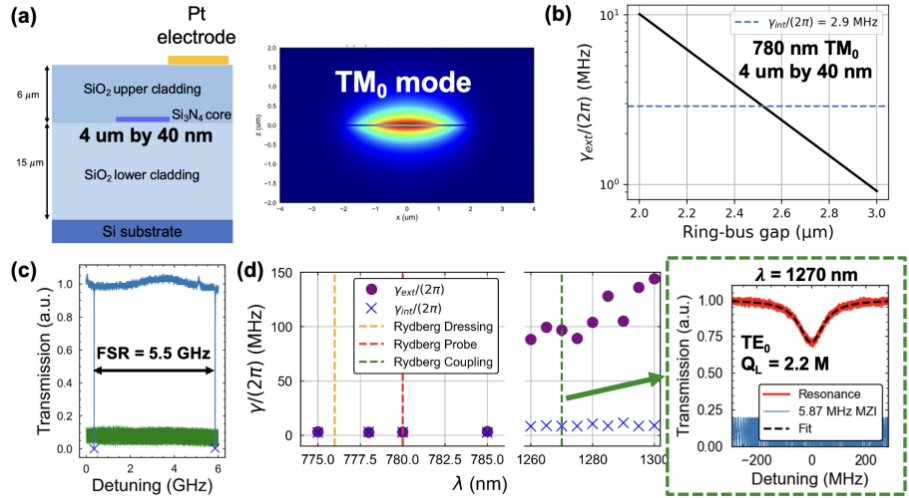

**Fig. S1**. Resonator design and characterization. (a) Cross-sectional schematic of the fabricated device showing the $Si_3N_4$ core layer, $SiO_2$ cladding thicknesses, and Pt heater used for tuning. The simulated $TM_0$ mode profile is also shown. (b) Simulated ring-bus gap sweep showing extracted external coupling rate $\gamma_{ext}$ at 780 nm, with dashed line marking the $\gamma_{int}$ internal loss rate at 2.5 um gap. (c) Measured free spectral range (FSR) of 5.5 GHz at 780 nm. (d) Extracted $\gamma_{ext}$ and $\gamma_{int}$ across 775–785 nm and 1260–1300 nm, including the rubidium Rydberg probe, dressing, and coupling wavelengths. The inset shows the resonance at 1270 nm ($TE_0$ mode), indicating a loaded Q of 2.2 million.

The waveguide supports both TE and TM modes and we design the ring-bus coupler to couple only the fundamental TM mode into the resonator using a gap of 2.5 um (Fig. S1(a,b)). The ring-bus coupling simulation is performed with a directional coupler model using coupled mode theory and mode solvers from Ansys Lumerical and Flexcompute Tidy3d (Fig. S1(b)). For each gap we simulate the ring-bus cross-coupling $\kappa^2$ and convert to a coupling (i.e. external) loss rate $\gamma_{ext} = c\kappa^2/(n_g L)$ where $n_g$ is the group index and $L$ is the resonator circumference, assuming $\kappa^2 \ll 1$. Critical coupling is achieved when $\gamma_{ext} = \gamma_{int}$ where $\gamma_{int} = c\alpha/n_g$ and $\alpha$ is the propagation loss. We measure the resonator quality factor and the free-spectral range (FSR) using a single-frequency 780 nm DBR laser containing an isolator (Fig. 2(a,b) and Fig. S1(c)). The DBR laser frequency is ramped by laser current control while monitoring the resonator transmission signal and the fringes of a fiber MZI of a calibrated FSR. Extracted loss rates across the 775–785 nm and 1260–1300 nm wavelength ranges are shown in Fig. S1(d), with rubidium Rydberg-relevant wavelengths indicated. We confirm coupling to the TM0 mode in the 775–785 nm region and TE0 mode in the 1260–1300 nm region. At 1270 nm, the measured $Q_L$= 2.2 M suggests that the resonator is suitable for PDH stabilization and for use as a transfer cavity for the Rydberg coupling laser, although in the present demonstration the 1270 nm beam was not passed through the cavity.

## 2. PDH laser stabilization feedback loop

**PDH lock schematic.** The 780 nm DBR laser is stabilized to a packaged photonic integrated resonator using the Pound-Drever-Hall (PDH) technique. The locking schematic is shown in Figure S2. An electro-optic phase modulator (EOPM) applies a 15.1 MHz phase modulation to the laser before light enters the integrated SiN cavity. The resonator has a loaded Q-factor of 69 million, corresponding to a total resonator cavity linewidth of 5.6 MHz. The reflected signal from the cavity is detected by a photodetector and the PDH error signal is digitally demodulated and filtered via the Red Pitaya digital lock-in system using the Linien locking tool [1]. Feedback is applied to the laser current using a Vescent laser current controller. PDH locking was performed with both the Red Pitaya and the Vescent D2-125 laser locking servo, and with both direct laser current and EOPM sideband generation.

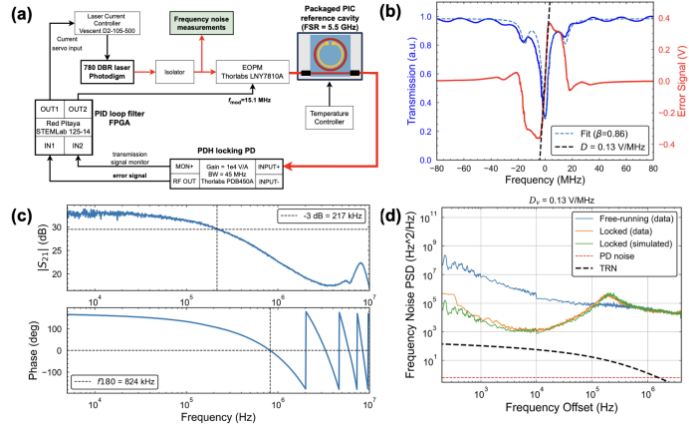

**Fig. S2**. PDH locking details. (a) Schematic of the Pound-Drever-Hall (PDH) laser frequency stabilization for locking the 780 DBR laser to the tunable integrated reference cavity. DBR: distributed Bragg reflector, EOPM: electro-optic phase modulator. (b) Transmission and error signal traces during laser frequency ramp used to extract the sideband modulation depth $\beta$ (unitless) and the error signal slope $D_\nu$ (units V/MHz). The 15.1 MHz sidebands are applied with the EOPM. (c) Network analyzer lock loop frequency response measurement during a weak lock using a Red Pitaya *pyrpl* network analyzer module. (d) Lock loop modeling with simulated frequency noise.

**PDH error signal and discriminator slope.** The CW laser field phase-modulated at the PDH dither modulation frequency generates sidebands given by [2]:

$$E_{in} = E_0 e^{i(\Delta\omega t + \beta \sin(\Omega t))} \approx E_0 e^{i\Delta\omega t}[J_0(\beta) + J_1(\beta)e^{i\Omega t} - J_1(\beta)e^{-i\Omega t}], \quad (S1)$$

where $\beta$ is the phase modulation depth, $\Omega/2\pi$ is the phase modulator modulation frequency, $J_n(x)$ is the $n$-th order Bessel function, and $\Delta\omega$ is the laser frequency detuning from the cavity resonance frequency. For an input power $P_{in} = |E_0|^2$, the optical power in the carrier is $P_c = J_0^2(\beta)P_{in}$ and the power in each first-order sideband is $P_s = J_1^2(\beta)P_{in}$. The modulated light passes through the resonator cavity which has a transfer function [3]

$$F(\Delta\omega) = \frac{i\Delta\omega + (\gamma_{int} - \gamma_{ext})/2}{i\Delta\omega + (\gamma_{int} + \gamma_{ext})/2}. \quad (S2)$$

The signal field is expressed by

$$E_{sig} = E_0 e^{i\Delta\omega t}[F(\Delta\omega)J_0(\beta) + F(\Delta\omega + \Omega)J_1(\beta)e^{i\Omega t} + F(\Delta\omega - \Omega)J_1(\beta)e^{-i\Omega t}] \quad (S3)$$

$$= E_0 e^{i\Delta\omega t}[FJ_0 + F_+ J_1 e^{i\Omega t} + F_- J_1 e^{-i\Omega t}].$$

Keeping only the $\Omega$ terms, the photodetected power at the transmission port of the add-thru resonator can be written as [3]

$$P_{sig} = |E_{sig}|^2 = 2 J_0 J_1 P_{in} |F F_+^* - F^* F_-| \cos(\Omega t + \phi_f). \quad (S4)$$

For sufficiently high modulation frequencies ($\Omega > \gamma_{tot}$) and near the cavity resonance ($\Delta\omega < \gamma_{tot}$) we have $|F(\Delta\omega)| \approx 1$ and

$$F F_+^* - F^* F_- \approx 2i \, \text{Im}[F] = \frac{2 \gamma_{ext} \Delta\omega}{|i\Delta\omega + \gamma_{tot}/2|^2} \approx \frac{8\gamma_{ext}\Delta\omega}{\gamma_{tot}^2}. \quad (S5)$$

After demodulation at $\Omega/2\pi$ we can extract the PDH error signal slope near the cavity resonance, in units of W/Hz:

$$\epsilon = \frac{P_{sig}}{\Delta\omega} = \frac{16 \, J_0 J_1 P_{in} \gamma_{ext}}{\gamma_{tot}^2}. \quad (S6)$$

For a critically coupled resonator ($\gamma_{ext} = \gamma_{int}$), the error signal slope is given by

$$D = \frac{P_{sig}}{\Delta f} = \frac{8 \sqrt{P_c P_s}}{\delta\nu}, \quad (S7)$$

where $P_c$ is the optical carrier power, $P_s$ is the sideband power, and $\delta\nu = \gamma_{tot}$ is the cavity total linewidth in Hz. In our experiments we extract the modulation depth from the transmission signal fitting to be $\beta = 0.86$ as shown in Fig. S2(b) corresponding to a sideband to carrier ratio of 0.2. For an input power $P_{in} = 80$ uW, photodetector TIA gain $10^4$ V/A, responsivity 0.5 A/W, and measured cavity linewidth 5.6 MHz, this corresponds to discriminator slope (expressed in V/Hz, hence $D_\nu$) of $D_\nu = 0.18$ V/MHz which is close to what we measured with our calibrated frequency axis error signal slope fit (0.13 V/MHz). Note that increasing $\beta$ beyond 1 does not significantly increase $D$, at which point the discriminator slope is ultimately limited by the cavity linewidth $\delta\nu$ [2]. In practice, the error signal slope can vary due to fluctuations in the photodetected power and the polarization of light into the PIC.

The discriminator slope $D_\nu$ can be used to estimate the noise $y_D$ due to the error signal generation:

$$S_{y,D} \approx \frac{S_D}{D^2}. \tag{S8}$$

The closed loop laser output FN from the contribution of the shot noise and PD noise can be found by using the expression for $D$:

$$S_{y,\text{shot}} = \frac{4h\nu P_s}{D^2} = \frac{(\delta\nu)^2 \, h\nu}{16 \, P_c}, \tag{S9}$$

$$S_{y,\text{PD}} = \frac{(\text{NEP})^2}{D^2}, \tag{S10}$$

where NEP is the noise equivalent power of the photodetector, in units W/$\sqrt{\text{Hz}}$. Using the measured discriminator slope, we calculate $S_{y,\text{shot}} = 9 \times 10^{-3}$ Hz$^2$/Hz and $S_{y,\text{PD}} = 0.6$ Hz$^2$/Hz, which are both below the resonator TRN.

**PDH lock loop modeling**. In PDH laser frequency stabilization, the resonator cavity acts as a frequency discriminator ("sensor") and a PID loop filter generates the control signal to the laser current controller to adjust the laser frequency. Below we derive the lock loop transfer function and use this to model the expected locked laser frequency noise.

The resonator cavity acts as the frequency reference $r$ that the laser frequency $y$ attempts to follow. The error signal in the reference tracking is therefore $e(t) = r(t) - y(t)$. In our system, we use the following transfer functions: $K(s)$ for the laser and laser current and temperature controller, $G(s)$ as the PID loop filter, and $D(s)$ as the resonator cavity frequency discriminator. In the frequency domain, the closed loop output signal can be expressed as [3]:

$$y(s) = K(s)G(s)D(s)e(s). \tag{S11}$$

Using our definition for the negative-feedback error signal and rearranging to isolate $y$,

$$y(s) = \frac{K(s)G(s)D(s)}{K(s)G(s)D(s) + 1} r(s). \tag{S12}$$

Here, the open loop transfer function can be defined as $L(s) = K(s)G(s)D(s)$. In practice, each block has a "disturbance" noise source, and the full output signal is

$$y(s) = K(s)\{G(s)[D(s)e(s) + y_D(s)] + y_G(s)\} + y_K(s), \tag{S13}$$

$$y(s) = \frac{L(s)}{L(s) + 1} r(s) + \frac{K(s)G(s)y_D(s) + K(s)y_G(s) + y_K(s)}{L(s) + 1}. \tag{S14}$$

The first term is the reference tracking, which has a limited bandwidth determined by the bandwidth of all the components and often limited by the laser itself. The second part is the in-loop residual noise that gets suppressed by the closed-loop feedback response of the PDH lock [3]. We use this to quantify how much frequency nose remains due to imperfect suppression. This includes the free-running noise of the laser $y_k$, the added noise in the error signal generation $y_D$ derived in the previous section to quantify noise due to the shot noise $S_{y,shot}$ and the photodetector noise $S_{y,PD}$, and $y_G$ the servo electronic noise. The parameters in this equation are summarized in Table S1.

**Table S1. Definitions for PDH lock loop transfer functions and noise modeling.**

| Parameter | Meaning |
|---|---|
| $K(s)$ | Laser and laser controller (e.g. current controller) |
| $G(s)$ | PID servo (e.g. Red Pitaya) |
| $D(s)$ | Resonator cavity frequency discriminator |
| $r(s)$ | Reference signal (intrinsic cavity noise) |
| $y_D(s)$ | Added noise from PDH error signal (e.g. PD and shot noise) |
| $y_G(s)$ | PID servo electronic noise (e.g. current noise) |
| $y_K(s)$ | Free-running laser noise |

The combined frequency response of the laser and laser controller can be modeled as:

$$K(s) = \frac{K_v}{1 + s/(2\pi f_{\text{rolloff}})}, \tag{S15}$$

where $K_v$ is the laser tuning strength which is measured to be $K_v$ = 280 MHz/V using a calibrated MZI measured during resonator Q measurements and monitoring the voltage input to the laser controller. The roll-off frequency $f_{\text{rolloff}} \approx$ 100 kHz is attributed to the small signal frequency modulation (FM) bandwidth of the semiconductor laser which is primarily limited by thermal effects [4]. The PID loop filter has the following transfer function:

$$G(s) = K_P + \frac{K_I}{s + 2\pi f_{I\,\text{lim}}} + K_D s, \tag{S16}$$

where we have included a bandwidth limit in the integrator term to reflect the FPGA's loop filter implementation limits. To characterize our PDH lock, we use the Red Pitaya PyRPL digital network analyzer (NA) module [5] to measure the small signal frequency response of the open-loop transfer function $L(s) = K(s)G(s)D(s)$. We overlay a small AC modulation at the current servo input while engaging a weak PDH lock to stay on the cavity resonance and to use the error signal slope as a frequency discriminator ("quasi-open-loop"). The measured response is shown in Fig. S2(c). We observe a -3 dB roll-off at 217 kHz, likely corresponding to the small-signal frequency modulation bandwidth of the laser. The 180° phase-lag frequency occurs at 824 kHz and ultimately sets the upper limit of the PDH locking bandwidth due to stability constraints such as the phase margin. We estimate the cable propagation delay of 50 ns corresponding to a 10 MHz 180° phase lag. The 320 ns propagation delay in the Red Pitaya Linien PDH locking application sets our 180° phase lag at 1.56 MHz.

Next, we apply our modeled PDH lock transfer function to the measured free-running DBR laser frequency noise to predict the locked laser performance. We implement the PDH loop model using the Python control repository [6], following the framework outlined in W. Wang et al. [7] including the resonator cavity low-pass filtering and lock loop delay. The simulated lock laser FN (Fig. S2(d), green trace) agrees well with the experimental data, confirming the model's accuracy in capturing the system dynamics.

As shown in Fig. 3(b) and Fig. S2(d), the PDH-locked DBR laser FN does not reach the cavity TRN limit. Based on our analysis we identify several possible factors that limit the PDH noise suppression in the 1-10 kHz frequency offset range. First, the finite feedback bandwidth set by the DBR current modulation response—estimated at 217 kHz from Fig. S2(c)—limits the achievable unity gain bandwidth (UGB) of the control loop. Second, the limited integral gain of the Red Pitaya servo, which we model using $f_{I,\text{lim}}$= 5 kHz. This reduces low-frequency loop gain and weakens noise suppression in the 1–10 kHz band. Lastly, the 5.6 MHz cavity linewidth imposes a low pass filter with corner frequency 2.8 MHz. This introduces a -20 dB/decade roll-off in the open-loop gain and contributes phase lag near and before the corner frequency, limiting the 180° phase lag point. Future work can explore the impact of the resonator linewidth (as described in [7]) on the ability of the PDH lock to reduce the FN to the TRN limit and to reach a lower $1/\pi$ reverse integration linewidth.

## 3. Cavity tuning and modulation

Here we characterize the effect of the cavity thermal tuner actuation on the PDH lock. We measured the optical frequency discriminator (OFD) frequency noise of the PDH locked laser for different DC heater powers applied to the cavity, demonstrating no noticeable change in the locked laser noise (Fig. S3(a)). For small-signal AC modulation, we applied a DC offset and a sinusoidal small signal oscillation to the cavity and measured the frequency response using the fiber unbalanced MZI (UMZI) as the frequency discriminator. This measurement was done while the lock was engaged. The setup used for measuring the response is shown in Fig. S4(b). For PDH locked measurements the circulator is replaced with a fiber splitter. We measure the 180° phase-lag frequency to be $f_{180°}$ = 15 kHz (Fig. S3(c)).

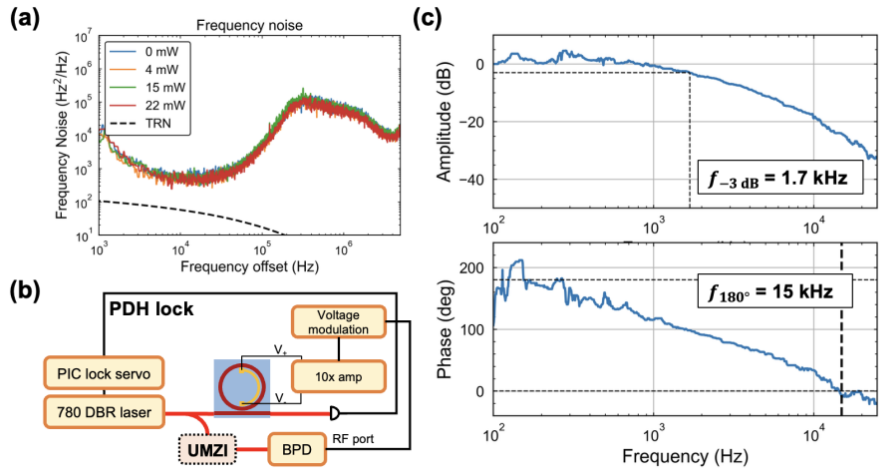

**Fig. S3**. Static and dynamic locked laser cavity tuning. (a) Frequency noise data of the PDH-locked laser for different tuner powers. (b) Pound-Drever-Hall (PDH) laser stabilization schematic to measure the small signal frequency response of cavity modulation for the PDH lock. The resonator is modulated with a waveform generator voltage source during frequency noise measurements. (c) Locked laser frequency response measurements by modulating the thermal tuner and measuring the RF port of the balanced photodetector signal from the UMZI.

## 4. Three-wavelength Rydberg electrometry

The three-wavelength Rydberg electrometry experiments were conducted at the University of Texas at Austin using a packaged, tunable resonator cavity device developed at UC Santa Barbara. The aim was to demonstrate how a single tunable integrated cavity can enable precise laser frequency stabilization for Rydberg atom-based electric field sensing. A three-wavelength excitation scheme was used to access high-lying Rydberg states. Compared to the two-wavelength Rydberg EIT approach (which often uses 480 nm and 780 nm lasers), three-wavelength approach can be achieved with relatively lower power near-IR and IR lasers and can achieve narrower EIT linewidths [8].

**Rydberg excitation scheme and energy levels.** The rubidium atomic ladder excitation scheme shown in Figure 5(a) of the main text consists of the following transitions: probe ($5S_{1/2} \rightarrow 5P_{3/2}$ at 780 nm), dressing ($5P_{3/2} \rightarrow 5D_{5/2}$ at ~776), and coupling ($5D_{5/2} \rightarrow nF_{7/2}$ at ~1270). The 1270 nm coupling laser is scanned to acquire the EIT signal, which appears as increased probe transmission when the Rydberg condition is met.

**Rydberg electrometry system.** The rubidium science cell used for these measurements is 4 inches long and heated to 37°C. The 780 nm and 1270 nm beams co-propagate, while the 776 nm beam is counter-propagating. All beams are collimated to a diameter of approximately 1 mm at the cell. Electric fields for Rydberg electrometry are applied using a horn antenna at 2.76 GHz, as shown in Figure 5(c) of the main text. Detection is performed via the 780 nm probe beam transmission and a dichroic mirror is used to filter out the co-propagating 1270 nm transmission. When the 1270 nm laser is on resonance with a Rydberg transition, EIT causes a measurable increase in the 780 nm probe transmission through the vapor cell.

**Laser stabilization with the tunable reference cavity.** We PDH stabilize the 780 nm Toptica ECDL probe laser to the integrated tunable cavity using laser current feedback, dither modulation frequency of 5 MHz, and a Toptica FALC laser locking module. Tuning across rubidium spectroscopy peaks is performed by thermally ramping the on-chip cavity thermal tuner. For long-term frequency stabilization to rubidium, we generate a secondary error signal either by applying an 11 kHz dither to the thermal tuner or by demodulating the existing 5 MHz modulation used for the PDH lock. The feedback to maintain lock to the rubidium $D_2$ spectroscopy peak is implemented using a Newport LB1005-S analog PID controller. For the 776 nm Toptica ECDL dressing laser, we use PDH sideband locking via an electro-optic phase modulator (EOPM). The EOPM applies a 1.1291 GHz modulation, enabling a lock to a sideband that matches the required 776 nm hyperfine transition frequency. This is necessary because the 5.5 GHz FSR of the integrated cavity does not directly coincide with the 776 nm transition. For all three locks we use the same photodetector at the transmission port of the add-thru resonator and demodulate the signal at the different modulation frequencies. A detailed schematic of the locking setup is provided in Figure 5(b) of the main text. In prior setups, Rydberg electrometry experiments required two separate vapor cells for laser stabilization: one dedicated to rubidium SAS at 780 nm, and a second "blue cell" used for detecting 420 nm fluorescence with a photomultiplier tube (PMT). In this configuration, the 780 and 776 nm lasers were independently stabilized using their respective reference cells before being routed into the science cell for spectroscopy. The tunable integrated cavity eliminates the need for the second reference cell entirely.

For comparison, we implemented a dual-laser stabilization scheme using a table-top ultra-low expansion (ULE) glass cavity with a four-bore design operated in vacuum using an ion pump. The EIT data for this is shown in the "QC Locked" trace in Figure 5(d) of the main text. In this configuration, the 780 nm and 776 nm lasers are independently PDH-locked to separate bores within the same cavity spacer, enabling simultaneous stabilization to a shared ultra-low-drift optical reference. This approach provides superior frequency noise suppression and lower intrinsic drift than the PIC-based stabilization system, owing to the high finesse of the ULE cavity. While the ULE cavity offers superior short-term frequency stability, rubidium spectroscopy provides better long-term referencing by anchoring the laser to an absolute atomic transition. However, the rubidium lock is subject to systematics such as probe beam power fluctuations which can limit its short-term performance. In our application, where the relevant timescales for Rydberg EIT measurements are on the order of minutes to hours, the combined Rb-disciplined PIC stabilization provides sufficient frequency stability without the complexity of a ULE cavity system. Future integration of the PIC cavity with on-chip Rb spectroscopy [9] can enable further system miniaturization.

**Frequency scanning and EIT trace monitoring.** The data presented in Figure 5(c) of the main text were obtained by scanning the 1270 nm external-cavity diode laser (ECDL) while monitoring the EIT signal for two cases: 1) 2.76 GHz radio-frequency (RF) turned on and 2) RF turned off. The raw measurements were taken using AC coupling in the oscilloscope; however, to remove the asymmetry in the pulse shape from the AC-coupled measurement, we applied the inverse of the AC-coupling filter to the data to recover the DC-coupled signal approximately. Thus, the estimated DC-coupled signal is presented in Figure 5c. The x-axis is converted from time to frequency detuning from the tuning strength calibration of the 1270 nm laser, which was calibrated at 118 MHz/V. As expected, we observe Autler-Townes splitting when the RF signal is turned on.

For the data in Figure 5(d) of the main text, we continuously monitored the EIT resonance height over a 20-minute interval with the RF field turned off. The rationale is to provide an approximate metric for the probe and dressing laser frequency stability. In these datasets, each point corresponds to the resonance height calculated from a complete scan across the Rydberg resonance, similar to the "RF off" trace in Figure 5(c). Similarly to Fig. 5(c), we applied the inverse AC coupling filter prior to computing the resonance height. Note that the EIT signal through the quad-cavity was sub-sampled after acquisition to match the other measurements. In Figure 5(d), we compare the results from three runs: 1) probe and dressing lasers unlocked, 2) probe and dressing lasers locked to rubidium stabilized PIC, and 3) probe and dressing lasers locked to quad cavity. In each case, we normalize the results to the resonance height at $t = 0$ to highlight the relative change in resonance height over the 20-minute window. The decrease in resonance height over time reflects cumulative effects such as laser frequency drift and reduced overlap with the optimal three-color resonance condition. As expected, in the cases where the probe and dressing lasers are locked to the PIC and quad-cavity, the EIT resonance height is maintained much better than in the case where the probe and dressing lasers are unlocked.

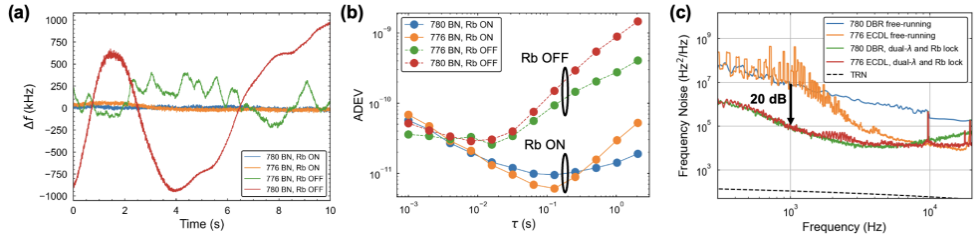

**Fig. S4**. Frequency stability measurements of the 780 nm and 776 nm transfer cavity locking using a beat-note (BN) with a ULE-cavity-stabilized, rubidium-disciplined optical frequency comb. (a) BN time series traces for the 780 nm DBR and 776 nm ECDL lasers both simultaneously locked to the PIC, with a moving average filter of 5 data points for the sampling rate 1000 samples/sec. The blue and orange traces correspond to dual-stage locking (PIC + Rb). The red and green traces show the unlocked Rb condition, with only the PIC locks of both lasers active. (b) Corresponding Allan deviation (ADEV) plots show improved frequency stability for both lasers when fully locked to Rb, with the stability of the 780 lock transferred to 776. Data were collected using a similar configuration as the UT Austin Rydberg electrometry setup, except here the 780 nm source is a higher-drift DBR instead of an ECDL. (c) OFD frequency noise measurements for the free-running 780 DBR and 776 ECDL and for the dual-wavelength and 780-rubidium-disciplined lock. The frequency noise at 1 kHz offset is reduced by 20 dB.

**Frequency stability of the multi-laser transfer cavity lock.** We evaluated the frequency stability of the multi-laser (780 and 776) transfer cavity lock at UC Santa Barbara by recording the beat-note (BN) of each laser with the ULE-cavity- and Rb-disciplined fiber frequency comb. The locking configuration followed that of Fig. 5(b) in the main text, with the exception that a higher-drift 780 nm DBR laser (Photodigm™) was used in place of a low-drift ECDL, and the 776 nm ECDL (Newport™ TLB-6730-P) was locked directly using PDH without sideband modulation. The BN measurements (Fig. S4(a)) for each laser were recorded sequentially because only one laser could be measured at a time. Although both lasers were locked to the PIC, the 776 nm ECDL exhibited frequency fluctuations of ±250 kHz over 10 s, whereas the 780 nm DBR laser showed larger excursions approaching ±1 MHz—likely due to intrinsic differences in free-running drift and lock quality between the two laser types. The corresponding Allan deviation (ADEV) traces are shown in Fig. S4b. With the PIC lock engaged, both lasers exhibit sub-$10^{-10}$ ADEV for averaging times of 1–10 ms. At longer averaging times (>10 ms), the addition of the Rb SAS lock to the 780 nm laser stabilizes the cavity resonance and transfers this enhanced long-term stability to the 776 nm laser, reducing its ADEV by approximately an order of magnitude at $\tau = 0.1$ s. Some residual differences in ADEV beyond 10 ms can be attributed to run-to-run variation in the spectroscopy lock and differing lock bandwidths for the two lasers. Simulations using the RydIQule repository [14] confirm that frequency drifts of ±250 kHz in the dual-stage-locked 780 nm and 776 nm lasers introduce negligible changes to the height or shape of the Rydberg EIT resonance. Instead, we attribute fluctuations in the observed EIT signal over time primarily to variations in laser power at the rubidium vapor cell and changes in cell temperature. We measure the frequency noise using the OFD system for the free-running and the multi-laser lock and observe a 20 dB reduction in the frequency noise at 1 kHz frequency offset (Fig. S4(c)). We attribute the noise reduction limits in part to the use of one photodetector in the lock loops for the two lasers and an interaction between the locks. The noise at these close-to-carrier frequencies for multiple lasers locked to the same cavity will be the subject of future work.

## References

1. B. Wiegand, B. Leykauf, R. Jördens, and M. Krutzik, "Linien: A versatile, user-friendly, open-source FPGA-based tool for frequency stabilization and spectroscopy parameter optimization," Review of Scientific Instruments **93**(6), 063001 (2022).

2. E. D. Black, "An introduction to Pound–Drever–Hall laser frequency stabilization," American Journal of Physics **69**(1), 79–87 (2000).
3. K. Liu, "Photonic integrated coil resonator stabilized narrow linewidth lasers and their applications," PhD Thesis, UC Santa Barbara (2024).
4. L. A. Coldren, S. W. Corzine, and M. L. Mashanovitch, *Diode Lasers and Photonic Integrated Circuits* (John Wiley & Sons, 2012).
5. L. Neuhaus, M. Croquette, R. Metzdorff, S. Chua, P.-E. Jacquet, A. Journeaux, A. Heidmann, T. Briant, T. Jacqmin, P.-F. Cohadon, and S. Deléglise, "Python Red Pitaya Lockbox (PyRPL): An open source software package for digital feedback control in quantum optics experiments," Review of Scientific Instruments **95**(3), 033003 (2024).
6. S. Fuller, B. Greiner, J. Moore, R. Murray, R. van Paassen, and R. Yorke, "The Python Control Systems Library (python-control)," in *2021 60th IEEE Conference on Decision and Control (CDC)* (2021), pp. 4875–4881.
7. W. Wang, S. Subhankar, and J. W. Britton, "A practical guide to feedback control for Pound-Drever-Hall laser linewidth narrowing," (2024).
8. S. M. Bohaichuk, F. Ripka, V. Venu, F. Christaller, C. Liu, M. Schmidt, H. Kübler, and J. P. Shaffer, "Three-photon Rydberg-atom-based radio-frequency sensing scheme with narrow linewidth," Phys. Rev. Appl. **20**(6), L061004 (2023).
9. M. T. Hummon, S. Kang, D. G. Bopp, Q. Li, D. A. Westly, S. Kim, C. Fredrick, S. A. Diddams, K. Srinivasan, V. Aksyuk, and J. E. Kitching, "Photonic chip for laser stabilization to an atomic vapor with $10^{-11}$ instability," Optica, OPTICA **5**(4), 443–449 (2018).